\documentclass[aps,twocolumn,superscriptaddress,amsmath,amssymb]{revtex4-1}
\usepackage{graphicx}
\usepackage{epsfig}
\usepackage{epstopdf}
\usepackage{bm}
\usepackage[usenames,dvipsnames]{color}

\begin{document}
\title{Topological electronic structure of YbMg$_2$Bi$_2$ and CaMg$_2$Bi$_2$}
\author{Asish K. Kundu}
\email{akundu@bnl.gov}
\affiliation{Condensed Matter Physics and Materials Science Division, Brookhaven National Laboratory, Upton, New York 11973, USA}
\author{Tufan Roy}
\affiliation{Research Institute of Electrical Communication, Tohoku University, Sendai 980-8577, Japan}
\author{Santanu Pakhira}
\affiliation{Ames Laboratory, Iowa State University, Ames, Iowa 50011, USA}
\author{Ze-Bin Wu}
\affiliation{Condensed Matter Physics and Materials Science Division, Brookhaven National Laboratory, Upton, New York 11973, USA}
\author{Masahito Tsujikawa}
\affiliation{Research Institute of Electrical Communication, Tohoku University, Sendai 980-8577, Japan}
\affiliation{Center for Spintronics Research Network, Tohoku University, Sendai 980-8577, Japan}
\author{Masafumi Shirai}
\affiliation{Research Institute of Electrical Communication, Tohoku University, Sendai 980-8577, Japan}
\affiliation{Center for Spintronics Research Network, Tohoku University, Sendai 980-8577, Japan}
\affiliation{Center for Science and Innovation in Spintronics, Core Research Cluster, Tohoku University, Sendai 980-8577, Japan}
\author{D. C. Johnston}
\affiliation{Ames Laboratory, Iowa State University, Ames, Iowa 50011, USA}
\affiliation{Department of Physics and Astronomy, Iowa State University, Ames, Iowa 50011, USA}
\author{Abhay N. Pasupathy}
\affiliation{Condensed Matter Physics and Materials Science Division, Brookhaven National Laboratory, Upton, New York 11973, USA}
\affiliation{Department of Physics, Columbia University, New York, NY, 10027, USA}
\author{Tonica Valla}
\affiliation{Condensed Matter Physics and Materials Science Division, Brookhaven National Laboratory, Upton, New York 11973, USA}
\date{\today}

\begin{abstract}
Zintl compounds have been extensively studied for their outstanding thermoelectric properties, but their electronic structure remains largely unexplored. Here, we present a detailed investigation of the electronic structure of the isostructural thermopower materials YbMg$_2$Bi$_2$ and CaMg$_2$Bi$_2$ using angle-resolved photoemission spectroscopy (ARPES) and density functional theory (DFT). The ARPES results show a significantly smaller Fermi surface and Fermi velocity in CaMg$_2$Bi$_2$ than in YbMg$_2$Bi$_2$. Our ARPES results also reveal that in the case of YbMg$_2$Bi$_2$, Yb-4$f$ states reside well below the Fermi level and likely have a negligible impact on transport properties. To properly model the position of 4$f$-states, as well as the overall electronic structure, a Hubbard $U$ at the Yb sites and spin-orbit coupling (SOC) have to be included in the DFT calculations. Interestingly, the theoretical results reveal that both materials belong to a $Z_2$ topological class and host robust topological surface states around $E_\mathrm {F}$. Due to the intrinsic hole doping, the topological states reside above the Fermi level, inaccessible by ARPES. Our results also suggest that in addition to SOC, vacancies and the resulting hole doping play an important role in the transport properties of these materials.

{\bf keywords:} Thermoelectric materials, Topological materials, Electronic structure, Surface state, Angle-resolved photoelectron spectroscopy, DFT.
\end{abstract}


\maketitle

\section{Introduction}

Intermetallic Zintl-phase compounds have been extensively studied because of their superior thermoelectric (TE) and magnetic properties and their possible use in various applications such as power generation, waste-heat conversion, and solid-state Peltier coolers \cite{snyder2011complex,disalvo1999thermoelectric,he2017advances,gascoin2005zintl,pakhira2021zero}. Recently, the research in this field has been stimulated by the discovery of nontrivial topological phases in various thermopower materials such as EuIn$_2$As$_2$ \cite{xu2019higher,regmi2020temperature}, EuSn$_2$As$_2$ \cite{li2019dirac,PhysRevB.104.174427}, and other CaAl$_2$Si$_2$-type compounds \cite{takane2021dirac,chang2019realization,marshall2021magnetic,kabir2019observation,Pakhira}. The possibility of a strain-induced topological phase transition is also predicted for $A$Mg$_2$Bi$_2$ ($A$ = Ca, Sr, Ba) compounds \cite{petrov2017effect}. Materials with nontrivial band topologies exhibit highly desirable properties, such as high carrier mobility, giant linear magnetoresistance, anomalous Hall effects, robust surface states, and high thermopower \cite{valla2012photoemission,li2016chiral,hoefer2015protective}.

\begin{figure*}[ht!]
\centering
\includegraphics[width=17cm]{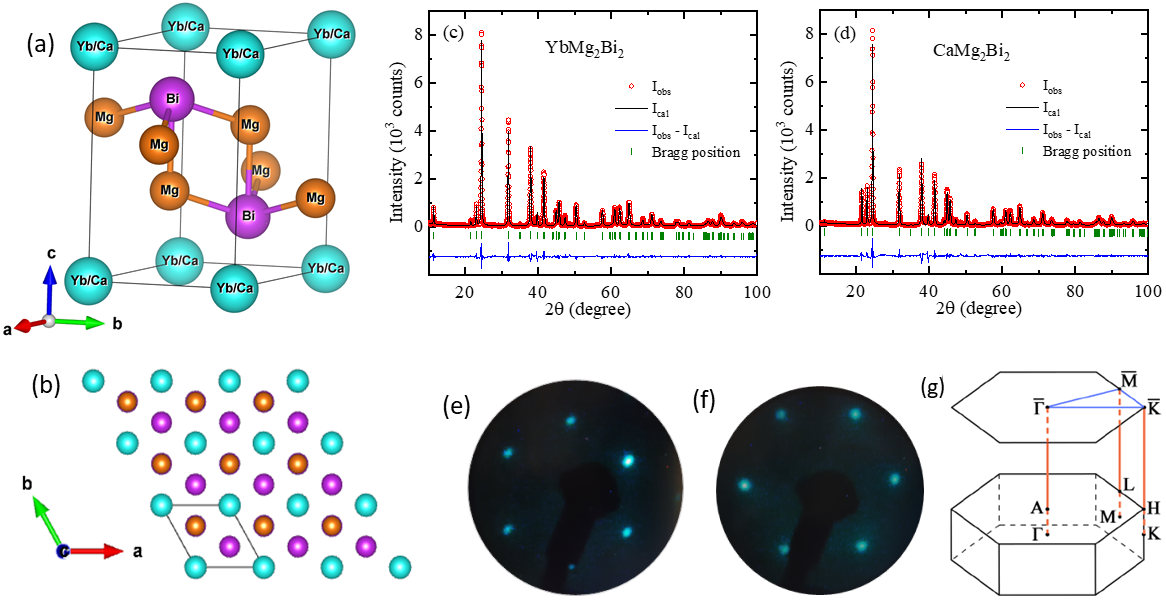}
\caption {Crystal structure of $A$Mg$_2$Bi$_2$ ($A$=Yb, Ca). (a) Unit cell of trigonal $A$Mg$_2$Bi$_2$. (b) Projection of the crystal structure onto the $a-b$ plane with a unit cell indicated by a rhombus. (c) and (d) XRD patterns of powdered YbMg$_2$Bi$_2$ and CaMg$_2$Bi$_2$ crystals, along with the refinement of the data using FULLPROF software. (e) and (f) Hexagonal LEED patterns of YbMg$_2$Bi$_2$ and CaMg$_2$Bi$_2$ single crystals obtained after cleaving the samples \textit{in-situ} under UHV, respectively. (g) Schematics of hexagonal bulk Brillouin zone and its surface projection (SBZ).}\label{Fig1}
\end{figure*}

Among Zintl families, $A$B$_2$X$_2$ ($A$ = Ca, Yb, Eu, Sr; $B$ = Zn, Mn, Cd, Mg; $X$ = Sb, Bi) type compounds gain more attention as many of them show high thermoelectric performance \cite{chen2018extraordinary,sun2017thermoelectric,shuai2016higher,wang2019experimental,wang2020enhanced,guo2020enhanced,zhang2016designing,
toberer2010electronic,sun2017thermoelectric,sun2017computational}. Recent works have demonstrated the coexistence of intrinsic magnetism, topological Dirac electronic states, and moderate thermopower efficiency in EuMg$_2$Bi$_2$ \cite{marshall2021magnetic,kabir2019observation,shuai2016higher,pakhira2021zero,pakhira2020magnetic}, which could provide a new playground to investigate the interplay between magnetism, topology, and thermoelectricity. Interestingly, it has been found that when the $A$-sites of these compounds contain rare-earth elements or are partially substituted by rare-earth elements, they show enhanced carrier mobility and carrier concentration than the alkaline-earth-containing compounds \cite{may2012thermoelectric,toberer2010electronic,may2011structure}. For example, going from CaMg$_2$Bi$_2$ to YbMg$_2$Bi$_2$, both the mobility and hole concentration are increased by a factor of $\sim$2 \cite{may2011structure}. This is quite unexpected, as for chemically-doped semiconductors, the mobility generally decreases with increasing carrier concentration \cite{wang2013metal}.

 The stronger hybridization between Bi and Yb/Eu was considered to be partially responsible for higher mobilities in Yb/Eu-based compounds \cite{shuai2016higher}. On the other hand, if the 4$f$ states were near the valence-band maximum, one could expect lower mobilities due to heavy bands and stronger scattering (impurity, electron-phonon, electron-magnon), exactly opposite to the experimental results \cite{may2011structure,shuai2016higher}. Flage {\it et al.} \cite{flage2010valence} have reported that in the prototype system YbZn$_2$Sb$_2$, the Yb-4$f$ states make a non-negligible contribution to the valence band edge. The exact role of heavy 4$f$ electrons in carrier concentrations and transport properties in rare-earth-based compounds is still unclear due to the lack of experimental studies.

To achieve a better TE performance,  high carrier mobility, heavy effective mass, and low lattice thermal conductivity are highly desirable \cite{snyder2011complex,shuai2016higher}. Various strategies have been proposed to achieve this goal, such as band convergence and resonant states for heavy effective mass \cite{pei2011convergence,pei2012thermopower,liu2012convergence}, band alignment to achieve high carrier mobility \cite{fu2014high,shuai2016higher} and introducing microstructural defects (point defects and nanostructures, etc.) and alloying to enhance phonon scattering \cite{xie2013beneficial,shuai2016higher,pei2012thermopower}. However, many of the theoretical calculations \cite{zhang2016designing,gong2016investigation,wang2021theoretical,zhou2020thermoelectric} do not take into account the effects of spin-orbit coupling (SOC) in the prediction of the transport properties. The effect of SOC could change the band degeneracy (which is related to carrier concentration), band hybridization, and band gap. Thus, the consideration of the effects of SOC is necessary for accurate predictions of new relevant materials and their associated properties. In addition to the more realistic modeling, the experimental measurement of the electronic structure is an irreplaceable component of studies of new materials.

Here, we report comprehensive experimental and theoretical studies of the  electronic structures of YbMg$_2$Bi$_2$ and CaMg$_2$Bi$_2$ using ARPES and first-principles calculations. Our results show the importance of inclusion of the SOC in describing these materials. We also show that the Yb-4$f$ states are far away from $E_\mathrm {F}$, implying that they do not play a role in transport. Our studies show that both materials are narrow band-gap topological insulators with topological surface states.

 \begin{figure*}[ht!]
\centering
\includegraphics[width=17cm]{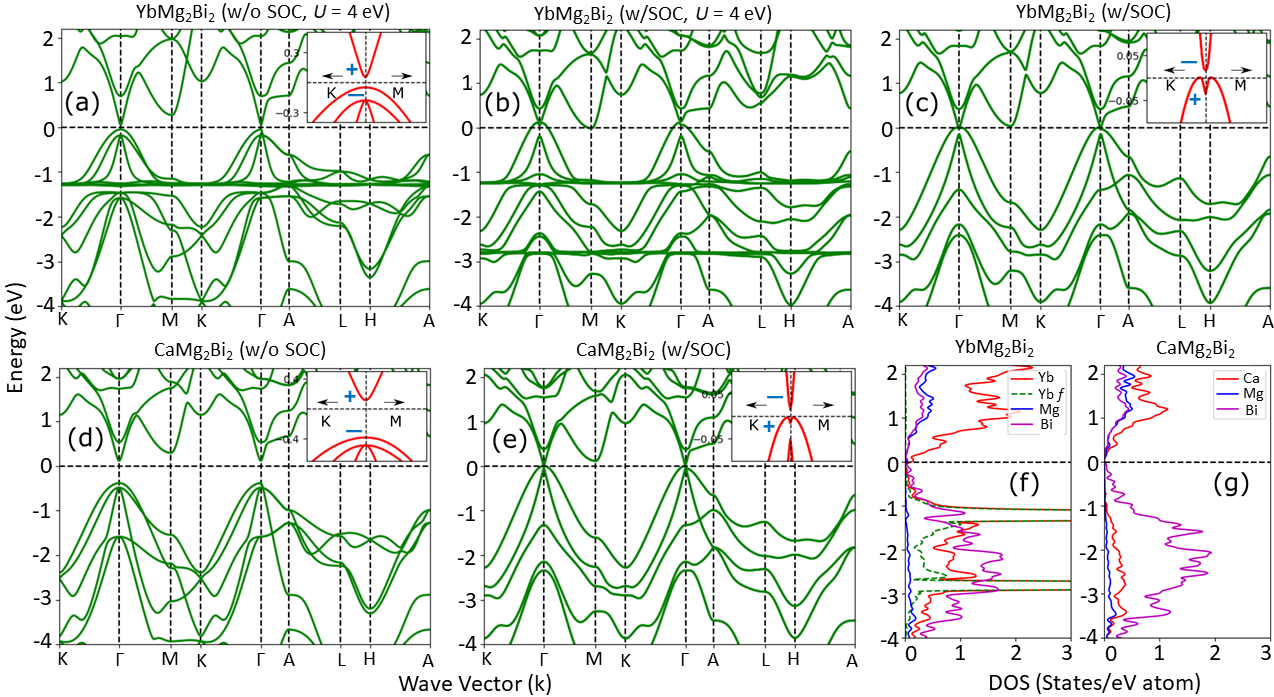}
\caption {Bulk band-structure calculations of $A$Mg$_2$Bi$_2$ crystals. (a) and (b) Band dispersions of YbMg$_2$Bi$_2$, without and with SOC, respectively. A Hubbard $U$ = 4 eV is included to treat the strongly-correlated Yb-$4f$ states. (c) Same as (b), but taking $4f$ electrons as core electrons. (d) and (e) Band dispersions of CaMg$_2$Bi$_2$ without and with SOC, respectively. The insets of (a), (c) and (d), (e) show the zoomed-in view of the fundamental band-gaps for the corresponding models. The parities of the topmost valence band and the first conduction band at the $\Gamma$ point are marked. (f) and (g) Partial density of states (PDOS) plots of YbMg$_2$Bi$_2$ (including $f$-electrons) and CaMg$_2$Bi$_2$, respectively.}\label{Fig2}
\end{figure*}

\section{Results and discussions}

\subsection{Structural details of samples}

Figure \ref{Fig1} shows the schematics of the crystal structure, x-ray diffraction (XRD) and the low-energy electron diffraction (LEED) patterns of the $A$Mg$_2$Bi$_2$ ($A$ = Yb, Ca) samples. Figures \ref{Fig1}(a) and 1(b) show the trigonal unit cell and the top views of the (001) surface (a-b plane) of the crystal structure, respectively. Using the Zintl concept, the trigonal structure of $A$Mg$_2$Bi$_2$ compounds can be viewed as polyanionic [Mg$_2$Bi$_2$]$^{2-}$ layers stacked along the c-axis and separated by the trigonal layers of $A^{2+}$. The room-temperature powder XRD patterns taken on crushed YbMg$_2$Bi$_2$ and CaMg$_2$Bi$_2$ single crystals are shown in Figs.~\ref{Fig1}(c) and 1(d), respectively. The experimental data fit well with the CaAl$_2$Si$_2$-type crystal structure having a trigonal lattice with space group $P\bar{3}m1$ (No. 164). The refined lattice parameters are $a = b =$ 4.7258(4) {\AA} and $c =$ 7.6453(14) {\AA}  for YbMg$_2$Bi$_2$ and $a = b =$ 4.7236(3) {\AA} and $c =$ 7.6512(10) {\AA} for CaMg$_2$Bi$_2$, in good agreement with the earlier-reported values \cite{may2011structure}. LEED patterns taken from in-situ UHV-cleaved YbMg$_2$Bi$_2$ and CaMg$_2$Bi$_2$ crystals are shown in Figs.~\ref{Fig1}(e) and (f), respectively. Hexagonal patterns are obtained for both samples, confirming that the cleaved surface is the (001) plane. The schematics of the hexagonal bulk Brillouin zone and its surface projection, {\it{i.e.}} surface Brillouin zone (SBZ), are shown in Fig.~\ref{Fig1}(g).

\subsection{Theoretical bulk electronic structures}

Figure~\ref{Fig2} shows the bulk electronic structure along various symmetry paths for $A$Mg$_2$Bi$_2$. Figures~\ref{Fig2}(a) and 2(b) illustrate the band dispersions of YbMg$_2$Bi$_2$, without and with spin-orbit coupling (SOC), respectively. A Hubbard $U$ parameter of 4 eV is included to describe the localized 4$f$ states of Yb and to match the observed splitting.
Multiple Dirac-like band crossings are observed (e.g., at K and between K and M symmetry points) in the absence of SOC; however, most of the Dirac points become gapped in the presence of SOC due to lifting of band degeneracy. The Yb-$4f$ derived flat band that was initially at about $-$1.3 eV [Fig.~\ref{Fig2}(a)] splits into two levels, $4f_{5/2}$ ($-$2.85 eV) and $4f_{7/2}$ levels ($-$1.25 eV) in the presence of SOC. We also see the strong hybridization between the $4f$ states and highly-dispersive bands when they cross each other. To better understand the role of $4f$ states, we have also treated the $4f$ electrons as core electrons and the resulting band structure is shown in Fig.~\ref{Fig2}(c). Note that the low-energy electronic structure is very similar to the one in which the 4$f$ electrons are considered as valence electrons [Fig.~\ref{Fig2}(b)], except for a slight downward shift of the hole-bands around the ${\Gamma}$ point. To understand whether this apparent energy shift is related to the intrinsic properties of the 4$f$ electrons, we performed the calculation considering higher $U$ values (see SI, Fig. S1). By increasing the $U$ value from 4 eV to 8 eV, we essentially move the 4$f$ states much deeper in energy, but still observe a similar low-energy electronic structure without any energy shift of the hole-bands. This suggests that the energy shift is related to inclusion of the external parameter $U$ in the DFT calculation. Without SOC a spectral gap of about 0.1 eV can be observed between the valence and conduction bands, and the system behaves like a semiconductor while it is nearly gapless in the presence of SOC. Interestingly, a closer look around the ${\Gamma}$ point [inset of Fig.~\ref{Fig2}(c)] reveals that a small energy gap of about $\sim$ 15 meV is still present and the shapes of the valence band maximum (VBM) and conduction band minimum (CBM) suggest that there might be a band inversion. The parity of the bands at $\Gamma$ also reverses when SOC is turned on, suggesting a band inversion scenario. Similar electronic structures have also been demonstrated in previous theoretical studies; however, they were conducted without SOC \cite{shuai2016higher}.
 \begin{figure*}[ht!]
\centering
\includegraphics[width=18cm]{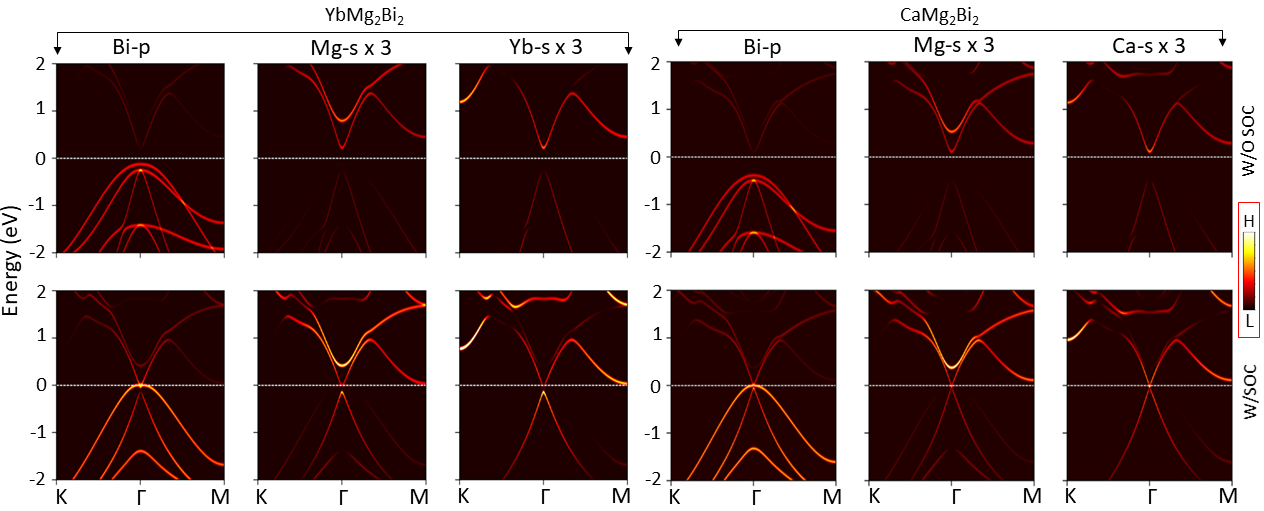}
\caption {Orbital-resolved band-structure of $A$Mg$_2$Bi$_2$. (a)--(c) and (d)--(f) Band dispersions of Bi-$p$, Mg-$s$ and Yb-$s$ orbitals without and with SOC for YbMg$_2$Bi$_2$, respectively. (g)--(i) and (j)--(l) Band dispersions of Bi-$p$, Mg-$s$ and Ca-$s$ orbitals without and with SOC for CaMg$_2$Bi$_2$, respectively. The spectral intensity of the Mg-$s$ and Yb-$s$ bands is multiplied by a factor of three to enhance the visibility.}\label{Fig3}
\end{figure*}

Similarly, Figs.~\ref{Fig2}(d) and 2(e) illustrate the electronic structure of CaMg$_2$Bi$_2$ without and with SOC, respectively. A spectral gap of about 0.5 eV can be seen between the VBM and CBM in the absence of SOC while the gap is extremely reduced when SOC is turned on. This is consistent with the previous reports \cite{shuai2016higher,sun2017computational}. The only major difference between these two systems is the absence of flat 4$f$ bands in the case of CaMg$_2$Bi$_2$. The band inversion is also reduced but is still present as shown in the inset of Fig.~\ref{Fig2}(e).

To correctly verify the nontrivial topology of these materials hinted by the shape of the valence/conduction bands near the zone center, we have calculated the $Z_2$ topological numbers using the Wilson loop (Wannier charge center) method \cite{yu2011equivalent} for the six time-reversal invariant momentum planes. The obtained results show $Z_2$ topological numbers $v_0$;$(v_1v_2v_3)$$=$1;(000) for both systems, which indicates that both materials are strong topological insulators. Here, we want to point out that although the low-energy electronic structure is very similar to the type-II nodal-line semimetal Mg$_3$Bi$_2$ \cite{chang2019realization}, the present materials are not nodal-line semimetals due to the presence of a band gap $\sim$ 0.10$-$0.5 eV without SOC. In contrast, nodal-line semimetals should show a conduction, and valence-band crossing in the absence of SOC.

Figures~\ref{Fig2}(f) and 2(g) show the partial densities of states (PDOS) of YbMg$_2$Bi$_2$ and CaMg$_2$Bi$_2$, respectively, in the presence of SOC. Both systems show very similar PDOS for Bi and Mg atoms, while it differs significantly for Ca and Yb atoms. Yb contributes more to the valence and conduction bands than Ca. The contribution of Yb-$f$ states is strongly localized around $-$1 eV, and it diminishes around the band edges. This implies a negligible role of $4f$ electrons in the transport properties in this material. In order to obtain a better understanding of the orbital character of the bands and their hybridization, orbital-resolved band dispersions are shown with and without SOC for YbMg$_2$Bi$_2$ [Figs.~\ref{Fig3}(a)--(f)] and CaMg$_2$Bi$_2$ [Figs.~\ref{Fig3}(g)--(l)] systems, respectively. The contributions from Ca/Yb-$s$, Bi-$p$, and Mg-$s$ are shown. It can be seen that the outer hole-like valence band has dominant Bi-$p$ character while the inner one has mixed Bi-$p$, Mg-$s$, and Ca-$s$/Yb-$s$ character. It appears that all orbitals are strongly hybridized with each other, both in the presence and absence of SOC. We note that some contributions of other orbitals such as Bi-$s$, Mg-$p$, Ca-$p$, Yb-$d$, and Yb-$p$ are also present around $E_\mathrm {F}$ (not shown).

\subsection{Comparison between the experimental and theoretical band dispersions: YbMg$_2$Bi$_2$}

 \begin{figure*}[ht!]
\centering
\includegraphics[width=18cm]{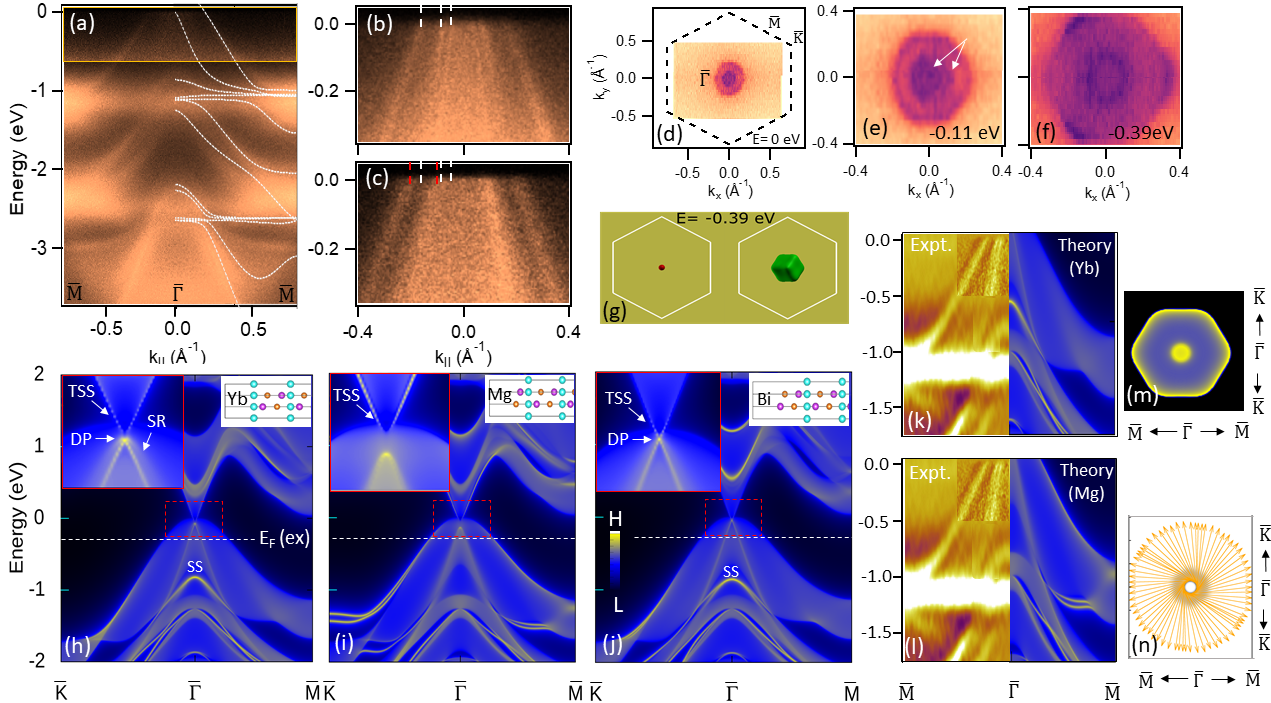}
\caption {Comparison of experimental and calculated electronic structure for YbMg$_2$Bi$_2$. (a) ARPES spectrum (from first SBZ) along the $\bar{M}-\bar{\Gamma}-\bar{M}$ line. The contrast at low binding energies is enhanced to improve visibility. Theoretical bulk bands (white dotted lines) along the ${\Gamma}-{M}$ path are superimposed on top. (b) and (c) ARPES spectra close to $E_\mathrm {F}$ around $\bar{\Gamma}$ from the first and second SBZ, respectively. The white and red vertical dashed lines indicate the $k_\mathrm {F}$ positions inside the first and second SBZ, respectively. (d) Fermi surface measured by ARPES (intensity integrated within $\pm$ 10 meV around $E_\mathrm {F}$). The SBZ is indicated by the dashed hexagon. (e) and (f) Constant-energy contours at $E=-$0.11 and $-$ 0.39 eV, respectively. Arrows in (e) indicate two nearly circular contours. (g) Calculated 3D constant-energy contour at $E=-$0.39 eV viewed perpendicular to the (001) surface. Spherical (left) and hexagonal (right) shaped contours are formed by the inner and outer hole-like bands at ${\Gamma}$. (h)--(j) Calculated surface state spectrum from a (001) semi-infinite slab of Yb-, Mg-, and Bi-terminated surfaces, respectively. Insets show the structure of different terminations (right) and zoomed-in spectra around $\bar{\Gamma}$ (left), within the dashed rectangles. The topological surface-state (TSS) is visible within the bulk gap. A surface-resonance band (SR) is marked in the inset of (h). The Dirac point (DP) is marked in the inset of (h) and (j). The horizontal dashed lines represent the positions of the experimental Fermi level. (k) and (l) Side-by-side comparison of the measured ARPES spectra with the calculated surface state spectra for Yb and Mg terminations, respectively. (m) Calculated constant-energy contours (at $E=-$0.3 eV) for the Mg-terminated surface. (n) Spin texture of TSS for the Mg-terminated surface.}\label{Fig4}
\end{figure*}

 Figure~\ref{Fig4} shows the comparison between the experimental and the calculated electronic structure of YbMg$_2$Bi$_2$. Panel~\ref{Fig4}(a) represents the measured ARPES spectrum along the ${\bar\Gamma}-\bar{M}$ line, which, at the photon energy used in the experiments, is very close to the bulk ${\Gamma}-{M}$ line. The calculated bulk electronic structure along the ${\Gamma}-{M}$ line is superimposed for comparison. To match the experimental dispersions, the theoretical spectrum is shifted up by $\sim$ 0.3 eV, implying that the measured sample is heavily hole-doped. According to the previous reports, intrinsic hole doping in these samples is predominantly due to $A$-site vacancies \cite{may2011structure,shuai2016higher}. In addition to numerous dispersing bands, two flat bands at energies $-$1.15 eV and $-$2.5 eV are visible in the ARPES spectrum. These flat bands are better resolved when probed with the He-II photon energy (40.8 eV) as shown in Fig. S2. These two flat bands arise from spin-orbit-split Yb-$4f$ states, matching nicely with the calculations when SOC and $U=$ 4 eV are taken into account. It is also visible that the dispersive bands and the Yb-$4f$ bands hybridize, in accord with our theoretical results. We note that the position of the Yb-$4f$ states varies among different Yb-compounds \cite{toberer2010electronic,antonov2002electronic}, implying that $U$ is not unique. In previous studies, a reasonable value of $U$ has been found to be 5.3 eV for YbZn$_2$Sb$_2$ \cite{toberer2010electronic} and 8 eV for YbB$_{12}$ \cite{antonov2002electronic}. High-resolution ARPES spectra in the vicinity of $E_\mathrm {F}$, near the zone center in the first and second SBZ, are shown in panels~\ref{Fig4}(b) and 4(c), respectively. Both spectra show linearly-dispersive hole-like bands crossing the Fermi level without any measurable renormalization, indicative of weak electron correlation in YbMg$_2$Bi$_2$. A visible change in the Fermi wave vector ($k_\mathrm {F}$) of both the inner and outer bands between the first and second SBZ originates from the difference in probed $k_z$ for the two spectra and points to their bulk electronic origin. This change of $k_z$ can be better realized from the momentum-dispersion curves (MDC) around $E_\mathrm {F}$ as shown in Fig. S3(c). The bulk origin of these states is in good agreement with the recent photon-energy-dependent ARPES studies on similar materials \cite{takane2021dirac,chang2019realization}. These linearly-dispersing bands form nearly-circular Fermi surfaces, Fig.~\ref{Fig4}(d). The constant-energy contours at $-$0.11 eV and $-$0.39 eV are also shown in Figs.~\ref{Fig4}(e) and 4(f), respectively. The inner two contours are very closely spaced and can only be resolved away from the Fermi level. The outer contour exhibits hexagonal warping that strengthens with increasing binding energy. The 3-fold symmetry of the outer contour also suggests its significant bulk-band character. Similar contours are obtained from the bulk electronic structure near the $\Gamma$ point of the 3D BZ, as shown in Fig.~\ref{Fig4}(g).

\begin{figure*}[ht!]
\centering
\includegraphics[width=16cm]{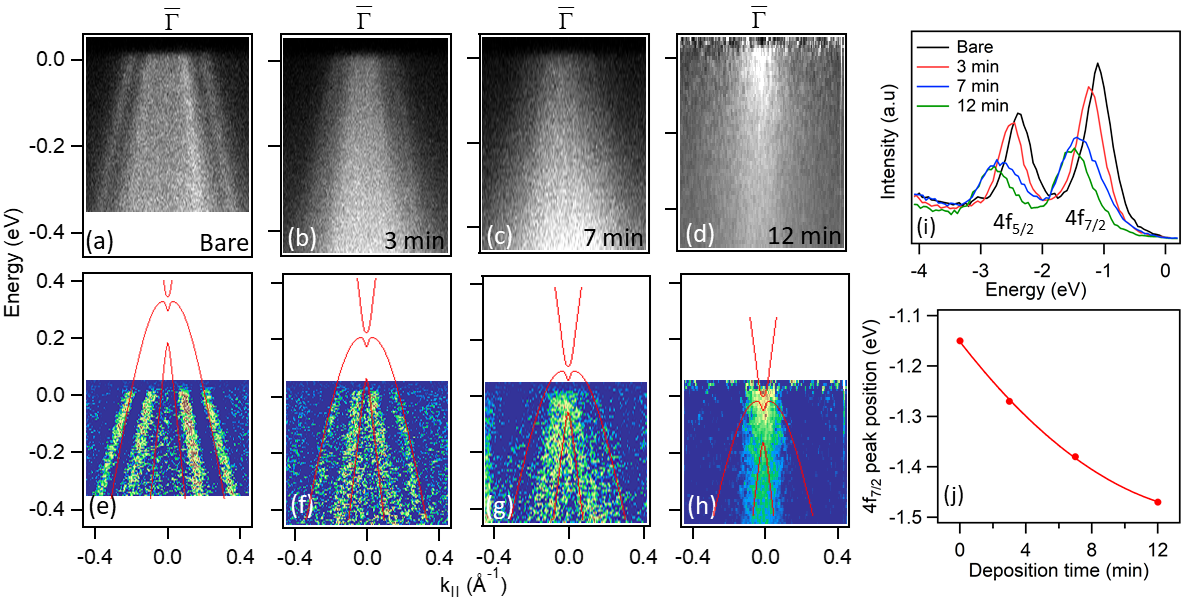}
\caption {Potassium deposition on YbMg$_2$Bi$_2$. (a)--(d) ARPES spectra for pristine YbMg$_2$Bi$_2$ and after 3, 7, and 12 min of potassium deposition, respectively. In (d), the ARPES intensity was divided by the Fermi-Dirac distribution function. (e)--(h) Second derivatives $d^2(I)/dk^2$ of the corresponding ARPES spectra along the momentum direction. (i) Corresponding energy distribution curves. (j) Energy of the 4$f_{7/2}$ peak obtained from (i).  Theoretical bulk bands (red curves) are overlaid on (e)--(h) after shifting the energy by amounts from (i).}\label{Fig5}
\end{figure*}

Since ARPES is a surface-sensitive probe and the bulk band-structure calculations imply the existence of topological surface states, we have also calculated the surface electronic structure for various possible terminations of the (001) surface. The Yb-4$f$ states are not included in these calculations for simplicity. Figures~\ref{Fig4}(h)--4(j) show the calculated surface electronic spectra from a (001) semi-infinite slab terminated by Yb, Mg, and Bi, respectively, where the enlarged views near the Fermi energy around $\bar{\Gamma}$ are shown in the left insets of each figure. The spectral brightness indicates the integrated charge density over the top six atomic layers, roughly equivalent to the ARPES probing depth.

 It is clear that a topological surface state (TSS), spanning the bulk band gap, is present on every termination. The Dirac point is buried inside the valence band, but should be exposed on Yb-and Bi-terminations. A similar TSS was also reported in other topological materials where the Dirac point lies inside the valence band, such as $p$-type Bi$_2$Te$_3$ and Mg$_3$Bi$_2$ \cite{hoefer2015protective,chang2019realization,zhou2019experimental}. The topological character of TSS is further supported by its helical spin-texture as shown in Fig.~\ref{Fig4}(n). Overall, the calculated spectra for different terminations are very similar, resembling the bulk band features. A few noticeable differences are: (1) in the case of Mg-termination, the outer state seems sharper, while for Yb and Bi terminations the inner state is more pronounced and (2) the Yb and Bi terminations show an intense surface state around  $\bar{\Gamma}$ at $-$0.85 eV, while the Mg termination does not. A direct side-by-side comparison of the ARPES spectra with the calculated surface state spectra [Figs.~\ref{Fig4}(k) and 4(l)] indicate somewhat better agreement with the Mg termination. However, the theoretical spectra for the Mg termination show only two states forming the Fermi surface [Fig.~\ref{Fig4}(m)], whereas three states are observed experimentally. On the other hand, the Yb terminations display an additional surface-resonance band (SR) [inset of Fig.~\ref{Fig4}(h)].

The coexistence of different terminations would probably give the best agreement with the observed spectra, as their contributions would be averaged over the macroscopic size of the excitation spot. To get further insight about the surface structure, we performed atomic force microscopy (AFM) measurements and the results are shown in Fig. S6. On the optically-flat surface regions, we observe different step heights, suggesting different surface terminations. The coexistence of different surface terminations has been observed for various materials such as Pt$_3$Te$_4$, YBa$_2$Cu$_4$O$_8$ and Bi$_4$Se$_3$ using micro/nano ARPES and scanning-tunneling microscopy (STM) \cite{fujii2021mitrofanovite,iwasawa2019buried,PhysRevB.88.081108}. Similar microscopic studies would be needed to disentangle the contributions from different terminations in the present materials.

\subsection{Potassium-deposited YbMg$_2$Bi$_2$}

\begin{figure*}[ht!]
\centering
\includegraphics[width=17cm]{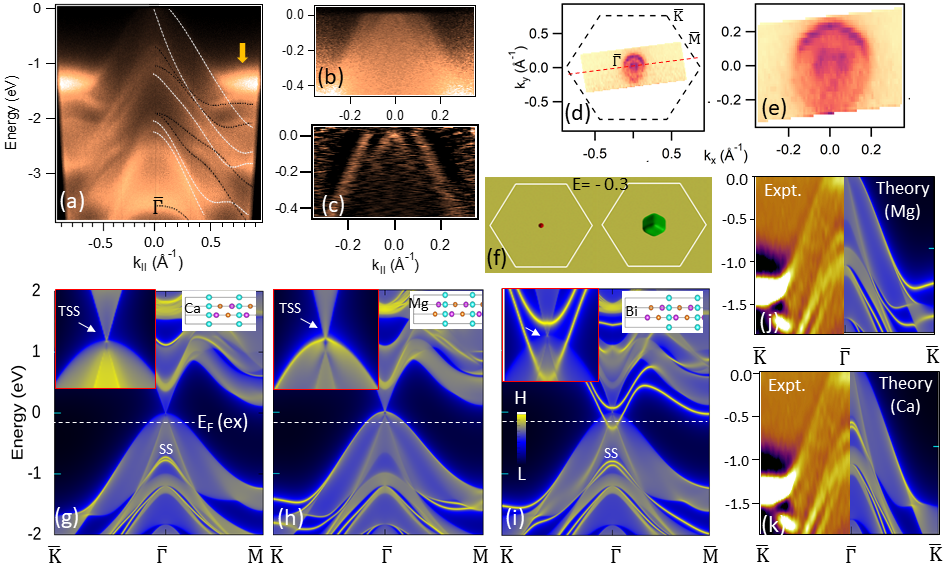}
\caption {Comparison of experimental and calculated electronic structure for  CaMg$_2$Bi$_2$. (a) ARPES spectrum along a line as shown by the red dashed line in (d). Theoretical bulk bands along ${\Gamma}-{K}$ (white dotted-lines) and ${A}-{H}$ (black dotted-lines) are superimposed after shifting them up by $\sim$ 0.15 eV. Theoretical bulk-bands capture most of the spectral features except for the state indicated by an arrow. (b) and (c) Low-energy ARPES spectrum and its second derivative [$d^2(I)/dk^2$], respectively. (d) and (e) Experimental Fermi surface and a constant energy contour at $E=-$0.10 eV, respectively. (f) Calculated bulk constant energy contours at $E=-$0.30 eV. Spherical (left) and hexagonal (right) surfaces are formed by the inner and outer hole-like bands at ${\Gamma}$. (g)--(i) Calculated surface-state spectra for Ca-, Mg-, and Bi-terminations. Insets show different terminations (right) and zoomed-in spectra around $\bar{\Gamma}$ (left) . The dotted lines represent the position of the experimental Fermi level. (j) and (k) Comparison of the ARPES spectra to the calculated surface-state spectra for Ca and Mg terminations, respectively.}\label{Fig6}
\end{figure*}

Due to high hole doping of pristine samples, the VBM, TSS and conduction band are unoccupied and cannot be probed in conventional ARPES. Therefore, we have performed {\it in-situ} electron-doping of cleaved surfaces by potassium deposition. The obtained results are shown in Fig.~\ref{Fig5}. Figure~\ref{Fig5}(a) shows the spectrum of the pristine YbMg$_2$Bi$_2$ surface in the second SBZ. The second SBZ is chosen because bands near $E_\mathrm {F}$ show a higher photoemission intensity. Figures~\ref{Fig5}(b)--(d) show the spectra after 3, 7, and 12 min of potassium deposition, respectively. It can be seen that the states shift towards higher binding energy and become blurred due to the increase in scattering rates. In the last sequence, an intensity (V-shaped-like feature) very close to $E_\mathrm {F}$ can be seen in Fig.~\ref{Fig5}(d), suggesting that the CBM just becomes occupied. We note that in Fig.~\ref{Fig5}(d), the ARPES intensity was divided by the Fermi-Dirac function to better resolve the spectral feature around $E_\mathrm {F}$. The raw ARPES data shown in SI (Fig. S4), are consistent with that conclusion. To better understand the change of electronic structure due to the potassium deposition, we have plotted the second derivatives of the spectra in Figs.~\ref{Fig5}(e)--(h) and the corresponding energy distribution curves (EDCs) in Fig.~\ref{Fig5}(i). The energy shift of the 4$f_{7/2}$ peak with potassium deposition time is shown in Fig.~\ref{Fig5}(j). A very similar energy shift is also observed for the Bi 5$d$ core-levels (Fig. S5), suggesting a nearly rigid-band-shift scenario. Thus, on top of the second derivative plots [Figs.~\ref{Fig5}(e)--(h)], the calculated bulk bands (red curves) are superimposed and shifted to match the experimental spectra by using the energies from Fig.~\ref{Fig5}(i). Even though there is a relatively good agreement between theory and experiment, we could not resolve the TSS and conduction band edge individually as the electronic states become too diffuse.

We note that the deposition of potassium on the surface alters the surface potential, causing band-bending in the relatively thick range from the surface ($\sim$ 10 -- 50 nm) \cite{bianchi2010coexistence}. Previous ARPES studies have shown that energy shifts of valence and conduction states are nearly equal in topological insulators \cite{kim2021absence,bianchi2010coexistence}. In case of YbMg$_2$Bi$_2$, the observed band structure shift is nearly-rigid, similar to that observed in other topological insulators \cite{kim2021absence,bianchi2010coexistence}.

\subsection{Comparison between experimental and theoretical band dispersions: CaMg$_2$Bi$_2$}

Figure~\ref{Fig6} shows the comparison between the experimental and the calculated electronic structure of CaMg$_2$Bi$_2$. Figure~\ref{Fig6}(a) represents the ARPES spectrum close to the $\bar{K}-\bar{\Gamma}-\bar{K}$ line. Theoretical electronic structures along the ${\Gamma}-{K}$ and ${A}-{H}$ lines are superimposed on Fig.~\ref{Fig6}(a). It is clear that most of the spectral features are well reproduced by the calculation, except for the states indicated by an arrow. For better visualization of the states in the vicinity of $E_\mathrm {F}$, a high-resolution ARPES spectrum is shown in Fig.~\ref{Fig6}(b) and its second derivative in Fig.~\ref{Fig6}(c). It is evident that two sets of hole-like bands cross the Fermi level. The FS and a constant energy contour (at $-$0.1 eV) formed by these bands are shown in Figs.~\ref{Fig6}(d) and 6(e), respectively. The shape of these contours is similar to that for YbMg$_2$Bi$_2$ and well reproduced by the theoretical calculation, as shown in Fig.~\ref{Fig6}(f).

Figures~\ref{Fig6}(g)--(i) show the calculated surface electronic spectra from a (001) semi-infinite slab terminated by Ca, Mg, and Bi, respectively. As in the case of YbMg$_2$Bi$_2$, a TSS (indicated by arrows) is present on every termination. In addition to the TSS, the Bi-terminated surface shows trivial surface states near the VBM that are absent on the Ca- and Mg-terminated surfaces. These surface states  originate from the unsaturated surface dangling bonds \cite{zhu2016electronic,tan2021double}. The absence of such states in the measured spectra [Figs.~\ref{Fig6}(a)--6(c)] indicates that the surface is not Bi-terminated. Direct comparisons with the calculated spectra for Ca and Mg terminations are shown in Figs.~\ref{Fig6}(j) and 6(k), respectively. Spectral features around $\bar{\Gamma}$ are in somewhat better agreement with the Ca-termination, while near $\bar{K}$ the agreement is better with the Mg termination. Therefore, the coexistence of Mg and Ca terminations would probably give the best agreement with the observed spectra. The AFM measurements (Fig. S7) showing different step heights support that scenario.

\begin{figure}[ht!]
\centering
\includegraphics[width=8.5cm]{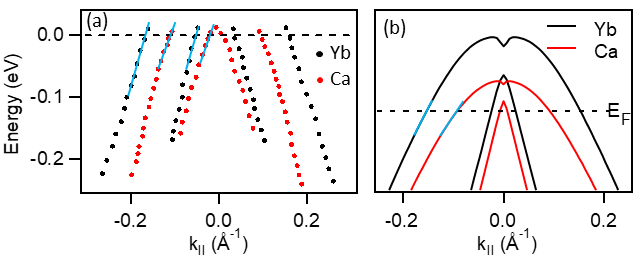}
\caption {Comparison of valence bands in YbMg$_2$Bi$_2$ and CaMg$_2$Bi$_2$. (a) MDC-derived bands of YbMg$_2$Bi$_2$ and CaMg$_2$Bi$_2$ near the Fermi level. The black (red) circles correspond to YbMg$_2$Bi$_2$  (CaMg$_2$Bi$_2$). Band slopes at $E_\mathrm {F}$ are highlighted by light-blue lines. (b) Shapes of the valence bands from our calculations. The CaMg$_2$Bi$_2$ spectra is shifted downward by 0.15 eV to reproduce the experimental features. Experimental $E_\mathrm {F}$ position is marked by dotted-line. }\label{Fig7}
\end{figure}

We note that according to the Zintl formalism both materials should be nominally charge balanced and predicted to be narrow-gap semiconductors. However, samples are usually highly {\it p}-doped, suggesting that holes are generated due to intrinsic defects \cite{may2011structure}. According to the previous Hall studies \cite{may2011structure,shuai2016higher}, the estimated values of carrier concentrations and mobilities are lower in CaMg$_2$Bi$_2$ than in the rare-earth-based compounds YbMg$_2$Bi$_2$ and EuMg$_2$Bi$_2$. From ARPES measurements it is also possible to estimate the carrier concentrations by measuring the volume of the FS. In Fig.~\ref{Fig7}(a), the MDC-derived bands crossing the Fermi level in YbMg$_2$Bi$_2$ and CaMg$_2$Bi$_2$ are plotted together. The Fermi wave vectors ($k_\mathrm {F}$) of both inner and outer bands are larger in YbMg$_2$Bi$_2$ than in CaMg$_2$Bi$_2$, implying higher carrier concentrations. By assuming an isotropic spherical shape of the two FS, formed by the two hole-like bands, we estimate the hole concentrations to be $\sim1.4\times$10$^{20}$ cm$^{-3}$ and $\sim5\times$10$^{19}$ cm$^{-3}$ for YbMg$_2$Bi$_2$ and CaMg$_2$Bi$_2$, respectively. The values estimated from the Hall coefficient are 4.6$\times$10$^{19}$ cm$^{-3}$ and 1.7$\times$10$^{19}$ cm$^{-3}$, respectively \cite{may2011structure}. Also, CaMg$_2$Bi$_2$ has a lower Fermi velocity than YbMg$_2$Bi$_2$. This behavior can be realized by putting the calculated valence bands of both materials on the same graph with their observed fillings [Fig.~\ref{Fig7}(b)].

Based on the shape and sharpness of the ARPES spectra, it seems that the effective masses and scattering rates are similar in these systems, which could lead to similar mobilities. However, the exact quantification of mobility is not possible from ARPES as the transport scattering rates are, in general, different from quasiparticle ones and the relationship between the effective mass and mobility for non-parabolic bands might be quite complex \cite{PhysRevLett.108.187001}. Moreover, the origin of higher mobilities in YbMg$_2$Bi$_2$ is still unclear and remains an open question.

We note that in predicting/interpreting  the thermoelectric properties of these classes of materials, some studies disregarded SOC \cite{zhang2016designing,gong2016investigation,wang2021theoretical,zhou2020thermoelectric}, leading to overestimated gaps, while the experiments and inclusion of SOC in calculations led to better agreement.

In summary, we have probed the electronic structure of the Zintl-phase isostructural analogs YbMg$_2$Bi$_2$ and CaMg$_2$Bi$_2$ using ARPES studies complemented with first-principles calculations. It is shown that the different carrier concentrations in these materials appear primarily due to a relative shift in $E_\mathrm {F}$ with respect to the valence band edge. Our results confirmed that the localized 4$f$ electrons do not play a considerable role in the low-energy electronic structure of YbMg$_2$Bi$_2$. We have also shown the importance of the SOC effect in these materials. Calculations suggest that these materials are narrow-band-gap topological insulators that support topological surface states with helical spin-texture. Due to the intrinsic hole doping, the actual $E_\mathrm {F}$ is located well inside the valence bands in these materials, which further restricts the effect of nontrivial band topology to be reflected in the transport properties. By tuning the $E_\mathrm {F}$ to the bulk band gap through chemical substitution or strain could lead to the observation of exotic surface transport properties such as Shubnikov-de Haas oscillations, opening the possibility for their use in spintronics applications.

\section*{METHODS}
\subsection*{Single crystal growth}
Single crystals of CaMg$_2$Bi$_2$ and YbMg$_2$Bi$_2$ were grown with the self-flux solution-growth technique. The high-purity starting materials Ca (99.999\%), Mg (99.98\%), Bi (99.9999\%) from Alfa Aesar and Yb from Ames Laboratory were used in nominal compositions CaMg$_4$Bi$_6$ and YbMg$_4$Bi$_6$. The elements were placed into alumina crucibles and sealed under $\approx$ 1/4 atm of high-purity inert Ar gas inside silica tubes. The assembly was then heated to 900 $^\circ$C at a rate of 50 $^\circ$C/h where the temperature was maintained for 12 hours. The tubes were then cooled to 850 $^\circ$C in 1 h followed by cooling to 750 $^\circ$C in 10 h and finally to 650 $^\circ$C in 24 h. Bulk three-dimensional-shaped single crystals with trigonal facets were obtained by removing the excess flux using a centrifuge at the last temperature.

\subsection*{Experimental details}
For the ARPES measurements, the samples were mounted on a sample holder and cleaved in-situ just before the ARPES measurements. The surface structure of the cleaved surface was checked using a low-energy electron diffraction (LEED) and an atomic-force microscope (AFM) (Digital Instruments, NanoScope III). The ARPES experiments were carried out at OASIS-laboratory at Brookhaven National Laboratory using a Scienta SES-R4000 electron spectrometer with monochromatized He I$_\alpha$ (21.22 eV) and He II$_\alpha$ (40.8 eV) radiation (VUV-5k) \cite{Kim2018a}. The total instrumental energy resolution was $\sim$ 10 meV for He I$_\alpha$ and 20 meV for He II$_\alpha$. The angular resolution was better than 0.15$^{\circ}$ and 0.4$^{\circ}$ along and perpendicular to the slit of the analyzer, respectively. Most of the data were taken at $\sim$ 20 K.
\subsection*{Theoretical details}
First-principles calculations have been carried out using Quantum ESPRESSO \cite{giannozzi2009quantum,giannozzi2017advanced,giannozzi2020quantum}. For the exchange and correlation energy/potential we used the PBEsol functional \cite{perdew2008restoring}. The projected-augmented-wave \cite{PhysRevB.50.17953} method has been used to represent the core electrons. All the calculations have been carried out using experimental lattice parameter, given in Section II.A. The cut-off energy for the plane waves and charge density were set to 60 Ry and 520 Ry, respectively. A $k$-mesh of 12$\times$12$\times$10 has been used for the Brillouin-zone integration. The electronic structure calculations have been carried out both with and without SOC. Pw2wannier interface and WANNIER90 were used for the construction of the first-principle tight binding Hamiltonian \cite{mostofi2014updated}. Surface-states spectra have been calculated using the WANNIERTOOLS package \cite{wu2018wanniertools}.

\section*{DATA AVAILABILITY}
The data that support the findings of this study are available from the corresponding author upon reasonable request.

\section*{ACKNOWLEDGEMENTS}
This work was supported by the U.S. Department of Energy, office of Basic Energy Sciences, Contract No. DE-SC0012704. The research at Ames was supported by the U.S. Department of Energy, Office of Basic Energy Sciences, Division of Materials Sciences and Engineering. Ames Laboratory is operated for the U.S. Department of Energy by Iowa State University under Contract No. DE-AC02-07CH11358. This work was also supported in part by CSRN, Tohoku University.

\section*{AUTHOR  CONTRIBUTIONS}
A.K.K designed the study, performed ARPES experiments and written the manuscript with a key contribution from T.R and S.P. A.K.K, and T.V analyzed the ARPES data. T.R performed the theoretical electronic structure calculation. S.P grew the bulk single-crystals and performed the physical characterization. Z.W performed the AFM measurements. T.V and D.C.J edited the manuscript. All the authors discussed the results and reviewed the manuscript. T.V made contributions to development of the OASIS facility used herein.

\section*{Materials and Correspondence}
Correspondence should be addressed to A.K.K. (akundu@bnl.gov).

\section*{COMPETING INTERESTS}
The authors declare no competing interests.

\end{document}